\documentstyle[sprocl]{article}

 \newcommand{\be}{\begin{equation}}
 \newcommand{\ee}{\end{equation}}
 \newcommand{\ba}{\begin{eqnarray}}
 \newcommand{\ea}{\end{eqnarray}}
 \def\entry#1#2{\vbox{\hbox to 92truept{\hrulefill}\break
                \hbox{\vrule\vbox to 20truept{
                \vfill
                \hbox to 92truept{\hfill\quad\small #1\quad\hfill}\break
                \vfill
                \hbox to 92truept{\hfill\quad\small #2\quad\hfill}
                \break\vfill
                \hbox to 92truept{\hrulefill}}\vrule}}}

 \def\arrwv#1#2{\vbox to 36truept{\vfill
                \hbox to 92truept{\put(48,0){\line(0,-1){10}}\hfill}\break
                \vfill
                \hbox to 92truept{\hfill\small #1\hfill}\break
                \vfill
                \hbox to 92truept{\hfill\small #2\hfill}\break
                \hbox to 92truept{%
                \put(48,0){\vector(0,-1){10}}\hfill}\break
                \vfill}}

 \def\arrwh#1#2{\vbox to 20truept{\vfill
                \hbox to 92truept{\small\hfill #1\hfill}\break
                \hbox to 92truept{\rightarrowfill}\break
                \hbox to 92truept{\small\hfill #2\hfill}\break\vfill}}

 \def\d{\partial}
 \def\Journal#1#2#3#4{{#1} {\bf #2}, #3 (#4)}

 \def\ANP{\em Ann.\ Physics (N.Y.)}

 \def\IJMPA{{\em Int.\ J.\ Mod.\ Phys.} A}
 \def\IJMPD{{\em Int.\ J.\ Mod.\ Phys.} D}
 \def\MPLA{{\em Mod.\ Phys.\ Lett.} A}
 
 \def\NPBP{{\em Nucl. Phys.} (Proc.\ Suppl.)}
 \def\PLB{{\em Phys. Lett.}  B}
 
 \def\PRD{{\em Phys. Rev.} D}
\begin{document}

\title{INTEGRABLE MODELS IN TWO-DIMENSIONAL DILATON GRAVITY}

\author{M.\ CAVAGLI\`A}

\address{Max-Planck-Institut f\"ur Gravitationsphysik,
Albert-Einstein-Institut,\\
Schlaatzweg 1, D-14473 Potsdam, Germany\\
E-mail: cavaglia@aei-potsdam.mpg.de}

\maketitle

\abstracts{We briefly present two-dimensional dilaton gravity from the
point of view of integrable systems.}

\noindent
Recently, two-dimensional dilaton gravity (DG) models have been
extensively investigated both from classical and quantum points of view
because of their connection to string theory, dimensional reduced models,
black holes, and gravitational collapse. (For short reviews and references
see e.g.\ \cite{Filippov,JSM}) In this talk I would like to illustrate
another important aspect of DG: its relation to integrable systems. 
DG is described by the action
\be
S[\phi,g_{\mu\nu}]=\int d^2x\sqrt{-g}[U(\phi)R(g)+V(\phi)+
W(\phi)(\nabla\phi)^2]+S_M[\phi,g_{\mu\nu},f_i]\,,\label{action-gen} 
\ee
where $U$, $V$, and $W$ are arbitrary functions of the dilaton, $R$ is the
two-dimensional Ricci scalar, and $S_M$ represents the contribution of
matter fields $f_i$ which include any field but the dilaton $\phi$ and the
graviton $g_{\mu\nu}$. Classically we may always choose $U(\phi)=\phi$ and
locally set $W(\phi)=0$ by a Weyl-rescaling of the metric so, for a given
$S_M$, Eq.\ (\ref{action-gen}) describes a family of models whose elements
are identified by the choice of the dilatonic potential. For instance,
when $S_M=0$ (pure DG)  $V(\phi)={const.}$ identifies the (pure) 
Callan-Giddings-Harvey-Strominger model (CGHS), $V(\phi)=\phi$ identifies
the Jackiw-Teitelboim model, and $V(\phi)=2/\sqrt{\phi}$ describes the
two-dimensional sector of the four-dimensional spherically-symmetric
Einstein gravity after having integrated on the two-sphere.

According to their integrability properties, DG models can be roughly
divided in three classes: {\it i)} {\em Completely Integrable Models},
i.e.\ models that can be expressed in terms of free fields by a canonical
transformation. Remarkable examples are the pure DG and the CGHS models;
\cite{Filippov,CGHS} {\it ii)} {\em Completely Solvable Models}, i.e.\
models that cannot be analytically solved in terms of free fields but
whose general solution is known. Two-dimensional effective generalized
theory of 2+1 cylindrical gravity minimally coupled to a massless scalar
field \cite{CavaPRD} and DG with constant or linear dilatonic potential
minimally coupled to massless Dirac fermions \cite{CFF,Strobl} belong to
this class; {\it iii)} {\em Partially Integrable Models}, i.e.\ models
that are integrable in a 0+1 dimensional sector only, namely after
reduction to a finite number of degrees of freedom. In this category we
find, for example, DG minimally coupled to massless Dirac fermions with
arbitrary potential \cite{CFF} and two-dimensional effective models
describing uncharged black $p$-branes in $N$ dimensions. \cite{pbrane}

Completely integrable models are of particular interest from the quantum
point of view. In this case one is able to quantize exactly the theory (in
the free-field representation) and, hopefully, to discuss quantization
subtleties and non-perturbative quantum effects. (See e.g.\ Refs.\
\cite{JSM,CGHS,Birkhoff} for a review of the state of the art of the CGHS
model.) Often these models can be used to describe black holes and/or
gravitational collapse, so the quantization program is worth exploring. In
the remaining part of this contribution I will briefly illustrate how far
one can go in the quantization program for the simple case of pure DG. 

An immediate consequence of the complete integrability of pure DG is that
both the metric and the dilaton can be expressed in terms of a D'Alembert
field and of a local integral of motion independent of the
coordinates.\cite{Filippov} So, using the gauge in which the free field is
one of the coordinates, one finds that all solutions depend on a single
coordinate.  This property is nothing else than the generalization of the
classical Birkhoff theorem.  (For spherically-symmetric Einstein gravity
the ``local integral of motion independent of the coordinates'' is just
the Schwarzschild mass.)  The reduction of the theory to a
finite-dimensional dynamical system signals that pure DG is actually a
topological theory.  Hence, the model can be quantized using two
alternative, a priori non-equivalent, approaches. In the first approach
the theory is quantized by first reducing it to a dynamical system with a
finite number of degrees of freedom, namely using first the Birkhoff
theorem and then the quantization algorithm.  Conversely, in the second
approach the theory is quantized in the full 1+1 sector and the
topological nature of the system must be recovered a posteriori ({\em
Quantum Birkhoff Theorem}): \cite{Birkhoff}
\be
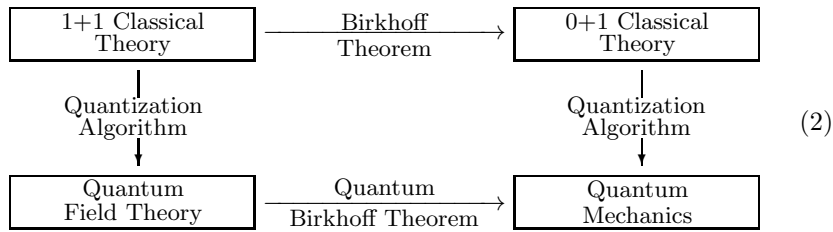
\begin{array}{rcc}
\entry{1+1 Classical}{Theory}
&\arrwh{Birkhoff}{Theorem}&\entry{0+1 Classical}{Theory}\\
\arrwv{Quantization}{Algorithm}&&\arrwv{Quantization}{Algorithm}\\
\entry{Quantum}{Field Theory}&\arrwh{Quantum}{Birkhoff Theorem}&
\entry{Quantum}{Mechanics}
\end{array}
\ee
A further ambiguity in the quantization procedure is related to gauge
invariance. Indeed, due to the coordinate reparametrization invariance of
the theory, the standard operator quantization of the system can be
implemented according to two different methods that are a priori
non-equivalent -- the {\em Dirac method} (quantization of the constraints
followed by gauge fixing) and the {\em reduced canonical method}
(classical gauge fixing followed by quantization in the reduced space). 
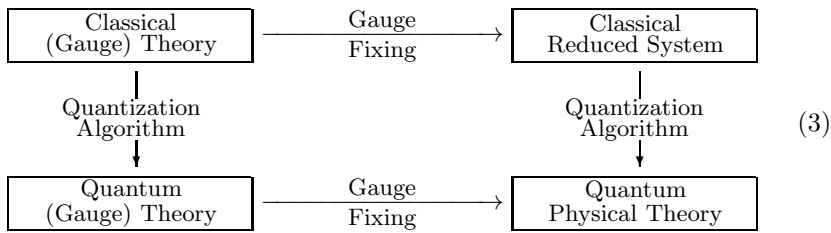
\be
\begin{array}{rcc}
\entry{Classical}{(Gauge) Theory}
&\arrwh{Gauge}{Fixing}&\entry{Classical}{Reduced System}\\
\arrwv{Quantization}{Algorithm}&&\arrwv{Quantization}{Algorithm}\\
\entry{Quantum}{(Gauge) Theory}&\arrwh{Gauge}{Fixing}&
\entry{Quantum}{Physical Theory}
\end{array}
\ee
It is really surprising that both diagrams can be closed and the
equivalence of the different approaches proved. Let me briefly illustrate
this point. 

In the 0+1 approach the closure of the second diagram is obvious. Since
the model is integrable we can solve the finite gauge transformations
generated by the (single) constraint $H=0$ and find a maximal set of
gauge-invariant canonical variables $\{M,P_M,H,T\}$. Then $T$ can be used
to fix the gauge since its transformation properties for the gauge
transformation imply that time defined by this variable covers once and
only once the symplectic manifold. The quantization becomes trivial and
both Dirac and reduced approaches lead to the same Hilbert space. This
program has been implemented in detail in Ref.\ \cite{bh} for the case of
spherically-symmetric Einstein gravity but can be easily generalized to an
arbitrary $V(\phi)$. The resulting Hilbert space is spanned by the
eigenvectors of the (gauge invariant) ``mass operator'' $M$.

Let us consider now the reduced quantization of the 1+1 theory. We may
find a canonical chart $\{M,\pi_M,\phi,\pi_\phi\}$ such that the ADM
super-Hamiltonian and super-momentum constraints read
\be
{\cal
H}=[N(\phi)-M]\pi_{\phi}\pi_M+[N(\phi)-M]^{-1}\phi'M'\,,~~~~
{\cal P}=-\phi'\pi_{\phi}-M'\pi_M\,,
\ee
where $'$ means differentiation w.r.t.\ the spatial coordinate $x_1$. 
Eventually, the canonical action must be complemented by a boundary term
at the spatial infinities of the form $S_\d=-\int
dx_0(M_+\alpha_++M_-\alpha_-)$ where $M_\pm\equiv M(x_0,x_1=\pm\infty)$
and $\alpha_\pm(x_0)$ parametrize the action at infinities. \cite{Kuchar}

Solving the constraints ($\pi_\phi=0$, $M'=0$) the effective Hamiltonian
coincides with the boundary term and the Hilbert space is spanned by the
eigenfunctions of $M\equiv M(x_0)$ with eigenvalue $m$. The Hilbert space
coincides with the Hilbert space obtained in the 0+1 approach. This proves
the equivalence of the 0+1 and 1+1 reduced methods of quantization. 

The equivalence between the 0+1 and the 1+1 Dirac methods can be proved
using a canonical transformation to free fields.  This has been done in
detail in Ref.\ \cite{Birkhoff} for the (pure) CGHS model. In this case an
explicit canonical transformation to free fields is known and the
quantization is carried out by use of the standard Gupta-Bleuler method. 
This is possible because the constraints can be linearized, due to
positivity conditions that are present in the model, leading to an
anomaly-free quantum theory. Again, the only gauge invariant operators are
the mass and its conjugate momentum and the vacuum must be labeled by the
eigenvalue of the mass operator. So there are infinite vacua, differing by
the eigenvalue of the mass, and one recovers the quantum mechanics of the
previous approaches.

\section*{Acknowledgements}

This work has been supported by a Human Capital and Mobility grant of the
European Union, contract no.\ ERBFMRX-CT96-0012.

\end{document}